# High quality a-axis outgrowth on c-axis $Y_{1-x}Ca_xBa_2Cu_3O_{7-\delta}$

G. Leibovitch[a], R. Beck and G. Deutscher

School of Physics and Astronomy, Raymond and Beverly Sackler Faculty of Exact Sciences, Tel Aviv University Ramat Aviv 69978, Israel

The large amplitude of the high Tc (HTS) superconducting gap is attractive for improved electronic applications. However, the study of such HTS cuprates has uncovered that unlike the s-wave order parameter of the low Tc, an angle dependent $d_{x^2-y^2}$ wave function is the dominant order parameter in such compounds. This symmetry causes low energy surface bound states, detrimental for applications, except at (100) oriented surfaces. It is therefore essential to have a smooth and well oriented surface of the crystallographic a-axis (100). In this work we present a study of an unconventional way to attain such surfaces in the form of a-axis outgrowth on a c-axis surface of sputtered $Y_{1-x}Ca_xBa_2Cu_3O_{7-\delta}$ thin film. The grains topography was tested using X-ray, SEM and AFM together with Point Contact and Tunnel Junctions measurements.

---

[a] email: guyguy@post.tau.ac.il

## Introduction

Tunneling spectroscopy has been a major tool for the study of LTSC[1], and an advance in the technology has allowed applications of Josephson effects[2]. The large value of the energy gap in the HTSC makes them very attractive to extend the range of Josephson applications. But control of HTSC junctions has proved to be a much more difficult task than that of LTSC junctions.

The best tunneling spectroscopy data so far has been achieved by STM, first on single crystals[3] and more recently on thin films[4,5]. STM has revealed the strong anisotropy in the Tunneling Density of States (TDOS) of the HTSC. The TDOS depends on the *local* orientation of the surface. If it is perpendicular to the Cu planes ((001), c-axis orientation), the TDOS gives the average in-plane DOS[3]. If it is parallel to the Cu planes, and parallel to one of the principal axis ((100), a-axis orientation), the TDOS is similar to that of a conventional superconductor, with a flat bottom at low bias[6]. If it is parallel to the Cu planes and at 45 degrees from the principal axis ((110) orientation), the TDOS is dominated by a peak at zero bias[4], called the Zero Bias Conductance Peak (ZBCP)[6]. These results are fully consistent with a d-wave symmetry of the order parameter[7], $\Delta_d = \Delta_0 \cos 2\Theta$, where $\Theta$ is an angle measured from the a-axis. In particular, the ZBCP seen in the (110) orientation reflects the existence of low energy surface bound states resulting from interference effects between lobes of the order parameter of opposite signs[8,9]. The detailed shape of the characteristic is governed both by the orientation of the surface and a barrier parameter Z, introduced by BTK[10]. For a perfectly transparent interface, Z = 0. This high sensitivity of the TDOS to the local surface orientation is one of the major obstacles towards HTSC Josephson applications. For such applications one would ideally like to use (100) oriented surfaces in order to have the full benefit of the large gap and of the conventional shape of the TDOS[6]. The $I_cR_N$ product should then approach the gap value, which in YBCO is close to 17 meV[11]. Unfortunately, the surface of a-axis oriented films is not flat over areas typically used for making planar junctions. As a result, ZBCPs are always seen in the characteristics in in-plane tunneling on YBCO films[12,13]. As a matter of fact, characteristics measured on (100) and (110) oriented films are rather similar[14]. Modeling of surface roughness has been successful in reproducing the observed I(V) characteristics[15]. Direct measurement of surface roughness on (100) and (110) YBCO oriented films gives typically values around 10 or more nm, more than sufficient to justify this modeling. The observed ZBCPs are a clear indication that the superconducting gap is strongly suppressed over a large fraction of the surface, the low energy surface states making such films inadequate for junctions' applications.

The purpose of the present contribution is to point out to a new route towards achieving flat a-axis oriented surfaces. It is based on the occurrence of the needle like outgrowth of a-axis oriented grains at the surface of c-axis oriented films[16,17]. Their existence is well known and one generally tries to minimize them because they tend to degrade such properties as the critical current density and the surface resistance at microwave frequencies. What we shall show is that the surface of these outgrowths is very flat and that their density can be promoted in a controlled way, to the point where they cover a large fraction of the surface. We show that the detrimental surface states can be substantially reduced in such films.

## Sample preparation and characterization

The growth method is similar to that of c-axis YBCO films on LaAlO3 substrate[18], but the growth is performed at a different temperature. Generally speaking, when the temperature is lowered, the density of a-axis grains increases up to a point of maximum beyond which it decreases. Further decrease of the temperature degrades the superconducting qualities of the film. Another parameter that influences the grains density is the percentage of Ca. We have repeated the growth procedure using 5, 10 and 20 % Ca. As shown In Fig.1, the higher is the Ca doping level, the higher is the grain density that can be reached and the higher the required temperature to obtain the maximum grain density. For 20%, 10% and 5% Ca, the optimum growth temperatures are 760, 720 and 700 degrees C respectively, and the maximum a-axis grains coverage are 90%, 80% and 60%.

Fig.2 shows a comparison of different a-axis outgrowth cover pictures for 10% and 20% Ca obtained at different temperatures taken by SEM. The clear difference between the different growth temperatures is seen on the 10% Ca as well as the coverage difference for different Ca concentrations.

The a-axis grains size varies with their surface coverage. In the case where the grains are remote and isolated, their length can reach a few microns. It goes down to 0.2-0.5 microns in the case of high coverage. The highest coverage is approximately as high as 90% of the surface and includes areas where the coverage is almost 100%. The medium coverage range (~50%) is usually characterized by a high inhomogeneity revealing areas of 20% as well as 70% coverage. The grains height is of the order of 250 A for a growth time of 1.5h.

Due to the close 3:1 (c:a axis) ratio of YBCO lattice parameter, the on plane orientation of the grains is in both of the original a and b directions of the c-axis growth. Both directions are equally preferred. This can be seen by the fact that the needles like shaped grains are directed in both of these directions. Their long side is oriented along their c-axis.

Fig3. presents an AFM picture with a typical grain profile showing the grain smoothness and vertical step. The roughness of the grain surface was found to be 12A rms, which is of the order of plus/minus 2 a-axis YBCO lattice parameters. The (100) parallelism was verified by X-ray Rocking-Curve analysis (Omega-Scan)[19] showing a remarkable planar quality with disorientation of the order of that of the LaAlO3 substrate. Compared to (100) thin films grown on STO substrates, the Full Width Half Maximum is typically reduced on the grains outgrowth from $0.9^o$ to $0.5^o$.

## Conductance characteristics

Junctions were prepared on the outgrowth films either by Point Contact (PC) or by using an Indium counter-electrode.

In the PC method, a gold wire cut with a sharp razor blade is mounted on a differential screw and delicately put in contact with the superconducting sample. The barrier parameter of the junction Z varies according to the surface contamination and degradation, due to exposure to air, and with the pressure applied on the Au tip - YBCO contact. Typical PC resistances are of the order of 10Ω to 100Ω, with a contact size in the range of a few to several nanometers. The In junctions were made by pressing a freshly cut In wire onto the sample, as described in ref 13,20. Their size is macroscopic, in the range of a fraction of a millimeter.

Characteristics of relatively high Z values PC contacts (having resistances of the order of hundreds of ohms) revealed in most of the measurements no ZBCP. In the macroscopic In Tunnel Junctions a ZBCP was observed but significantly smaller than the ZBCP observed on typical (100) film grown on STO or LSGO substrates. Fig.4 shows a comparison between characteristics measured on a regular a-axis film and on an outgrowth film. In the regular a-axis film, the ZBCP is larger than the Gap Like Feature (GLF) seen around 15 meV. By contrast, in the outgrowth film the ZBCP is much diminished, the major feature is now the gap. However, a quantitative fit to theory assuming a pure (100) orientation is still not possible.

Fig.5 shows a low Z PC characteristic obtained on an outgrowth 10% Ca doped YBCO ($T_c$ = 75K), with a fit to theory. The fit parameters were the $\Delta_{x^2-y^2}$ gap amplitude, a possible additional imaginary component is or $id_{xy}$, the barrier strength Z and the smearing parameter $\Gamma$[9]. In view of the relatively low Z value required for the fit (Z ≈ 0.7), the tunneling cone width was taken as 180°. The gap value was obtained as 18 meV, giving a strong coupling parameter ($2\Delta/kT_c$) = 5.8. The margin of uncertainty on this value is small (1.8meV), in view of the small value of $\Gamma$ necessary to obtain a good fit. The quality of the fit could be improved with a small additional $id_{xy}$ (Fig. 5(a)), having a value of 1.2 meV. This value is in agreement with previous estimates in overdoped samples[21]. It was also possible to fit the data with an additional is order parameter.

However, this fit required a smaller gap, a high value of $\Gamma$, of more than 5 meV, and a very large is component order parameter or a larger Z (close to 1), together with a narrow tunneling cone. This set of parameters looks less likely. It would also be in disagreement with previous results on similar Z value juncitons[22]. We underline that the value of the gamma parameter used in the d + id fit is lower than that usually needed for similar Z contacts[23]. This suggests that PC junctions obtained on outgrowth films are more ideal than those obtained on regular a-axis films.

The difference between the improved but non-ideal behavior of macroscopic In junctions, and the almost ideal behavior of PC junctions is easily explained by their different sizes. The size of the PC tip is smaller than that of the typical outgrowth: it sees an almost ideal (100) surface. The size of an In junction is much larger that that of an outgrowth, and can still see different facet orientations.

**Conclusions**

In summary, we have grown thin films of YBCO doped with different concentrations of Ca. By changing the temperature of the growth we have managed to control the well known a-axis outgrowth of the c-axis films. We have reached the level <u>at</u> which one can say that the surface is mainly high quality a-axis oriented. The grains quality was tested by AFM and X-ray. Results of PC and In junctions support the good surface orientation, which appears substantially better than that of regular a-axis films. We believe that further manipulation of the outgrowth will prove to be applicable for the desired junctions on (100) surfaces of HTSC that are in need to take full advantage of the large gap of these superconductors.

This work was supported by the Israel Science Foundation, by the Heinrich Hertz Minerva Center for High Temperature Superconductivity and by the Oren Family Chair for Experimental Solid State Physics. We would like to thank Amir Kohen for fruitful discussions.

**Figures**

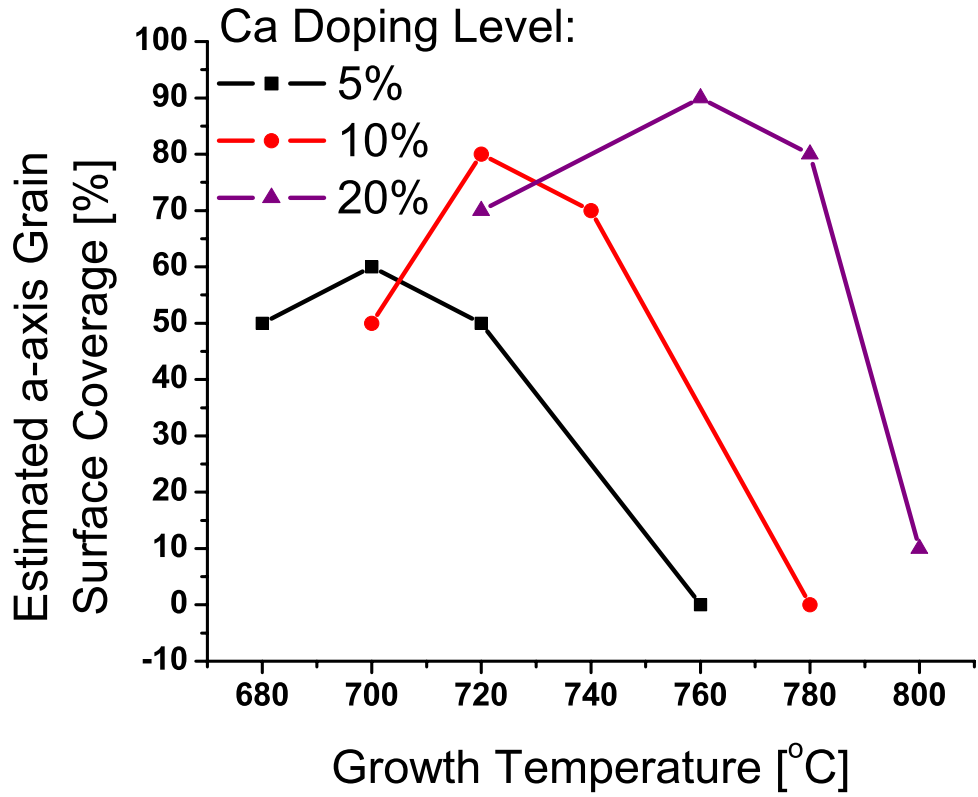

Fig1. a-axis grains surface coverage for different Ca doping levels. Maximum coverage and the corresponding growth temperature are reduced with reducing the Ca doping level.

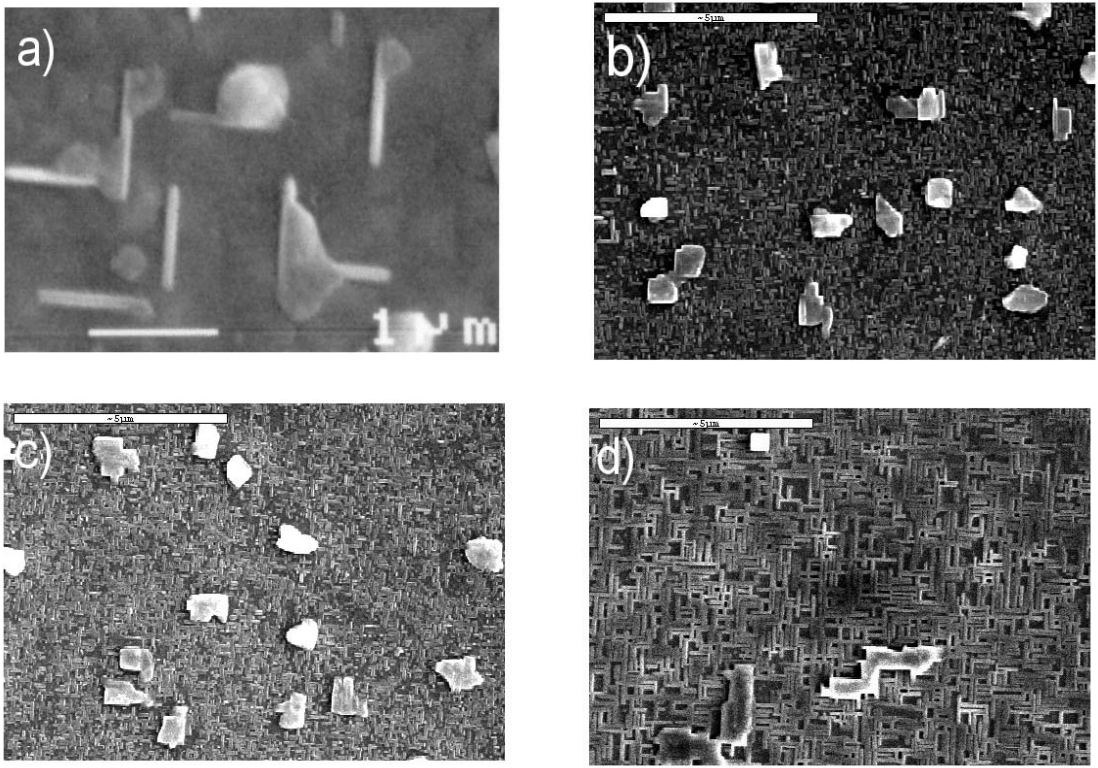

**Fig 2. Comparison between different a-axis grain surface coverage.**

**a) Scale line indicated is 1μ (5μ in b, c and d). The long needle like grains known outgrowth[16,17] .b) 10% Ca doped film growth temperature of 740K. c) 10% Ca doped film growth temperature of 720K. higher grain surface coverage. d) 20% Ca doped film growth temperature of 760K. high surface coverage revealing areas of 100% coverage.**

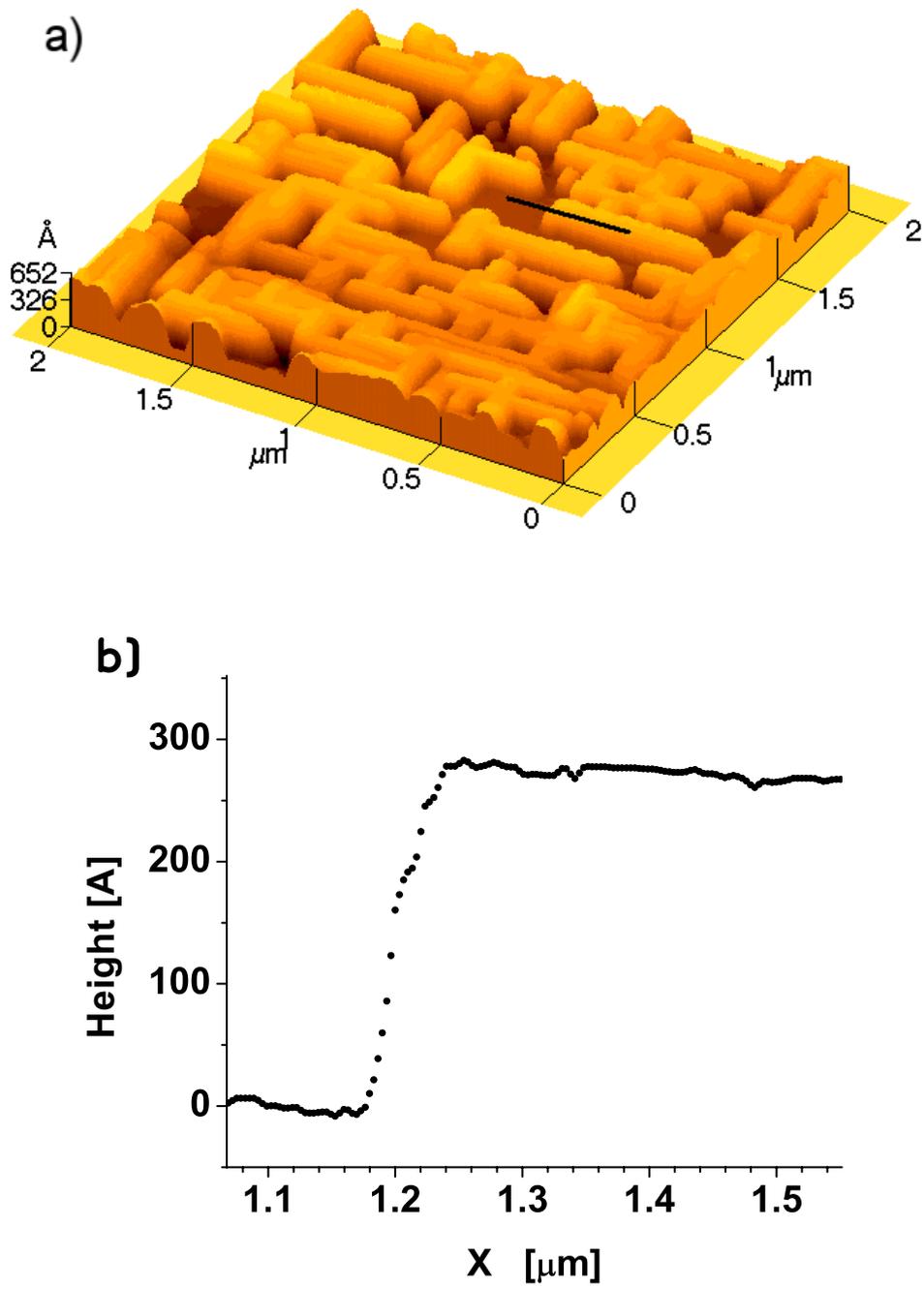

**Fig 3. a) 3D AFM pictures of a 20% Ca doped YBCO. b) Height measurement by AFM of the black line in 2(a). Presenting a sharp step followed by very small surface roughness.**

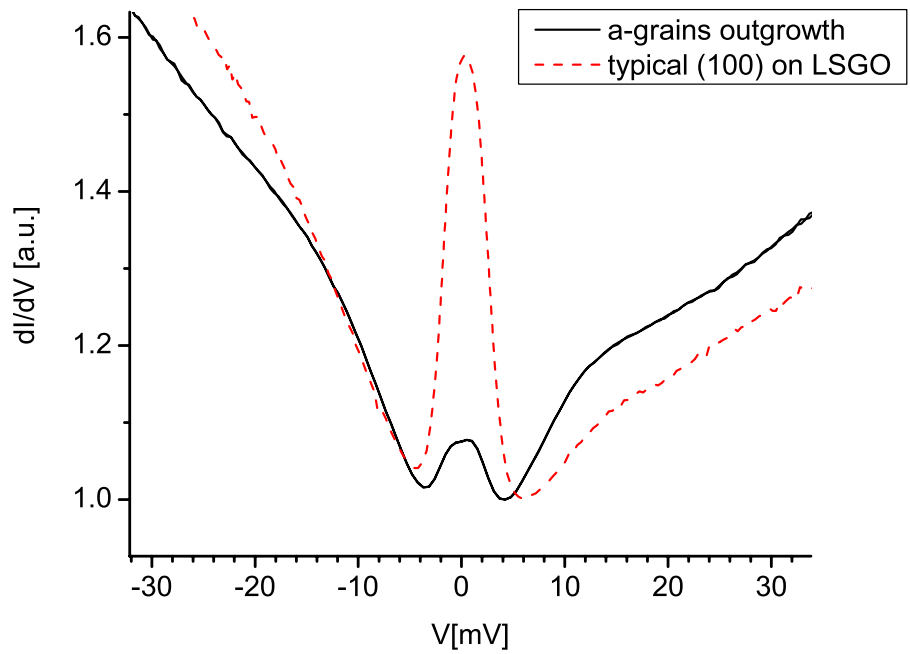

**Fig 4. Comparison of (100) In tunnel junctions. Dotted (red) line is a (100) YBCO-In Tunnel junction spectra with a strong ZBCP due to surface roughness. Continuous (Black) line is a junction on a 20% Ca doped sample with a high coverage of a-axis grains outgrowth.**

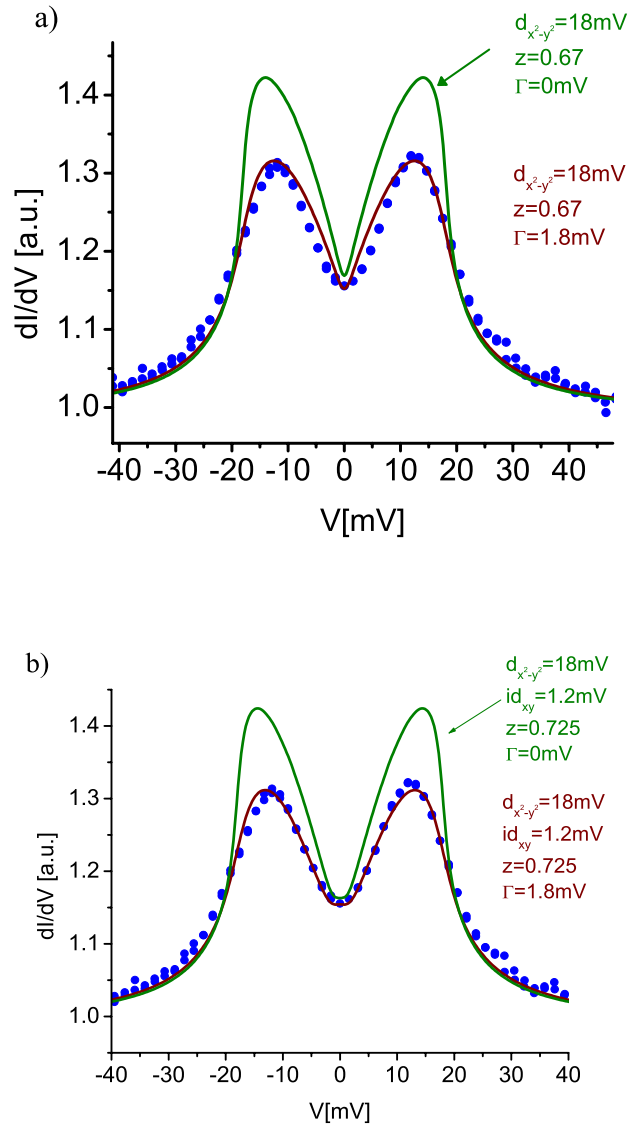

**Fig 5. Fitting of the conductance of a low Z PC on a 10% Ca doped sample to theory (ref.21). ( T=4.2K).**

**a) Fit using only a pure $d_{x^2-y^2}$ order parameter.  b) Improved fit using an additional subdominant order parameter $id_{xy}$.**

[1] W. L. McMillan and J. M. Rowell, in *superconductivity*, edited by R D Parks, Vol. **1**, Chap. 11, p.561.

[2] B. D. Josephson, in *superconductivity*, edited by R. D. Parks, Vol. **1**, Chap. 9, p.423.

[3] O. Fischer, C Renner, and M Aprile, in *The Gap Symmetry and Fluctuations in High-Tc superconductors*, edited by J. Bok et al. (Plenum press, Ney-York, 1998), Vol. **1**, p.487.

[4] A. Sharoni et al, Phys. Rev. B **65**, 134526 (2002).

[5] A. Sharoni, G. Leibovitch, A. Kohen, R. Beck, G. Deutscher, G. Koren, and O. Millo, Europhys. Lett. **62**, 883 (2003).

[6] A. Sharoni et al, Europhys. Lett. **54**, 675 (2001).

[7] C. C. Tsuei and J. R. Kirtley, Rev. Mod. Phys. **72**, 969 (2000).

[8] Cjia-Ren Hu, Phys. Rev. Lett. **72**, 1526 (1994).

[9] Y. Tanaka and S. Kashiwaya, Phys. Rev. Lett. **74**, 3451 (1995).

[10] G. E. Blonder, M. Tinkham, and T. M. Klapwijk, Phys. Rev. B **25**, 4515 (1982).

[11] Y. Dagan et al, Phys. Rev. B **62**, 146 (2000).

[12] M. Covington et al, Phys. Rev. Lett. **79**, 277 (1997).

[13] R. Krupke and G. Deutscher, Phys. Rev. Lett. **83**, 4634 (1999).

[14] R. Beck, A. Kohen, G. Leibovitch, H. Castro, and G. Deutscher, J. Low Temp. Phys. **131**, 445 (2003).

[15] M. Fogelstrom et al, Phys. Rev. Lett. **79**, 281 (1997).

[16] A. Kohen, PhD Thesis, Tel-Aviv University, Isral (2003).

[17] C. C. Chang et al, Appl. Phys. Lett. **57**, 1814 (1990).

[18] Y. Dagan et al, Phys. Rev. B **61**, 7012 (2000).

[19] R. Krupke, PhD Thesis, Tel-Aviv university (2003).

[20] Y. Dagan, Phys. Rev. Lett. **87**, 17704-1 (2001).

[21] G. Deutscher et al, Physica C **341-348**, 1629 (2000).

[22] A. Kohen, G. Leibovitch, and G. Deutscher, Phys. Rev. Lett. **90**, 207005 (2003).